\documentclass[twocolumn, trackchanges]{aastex63}

\submitjournal{ApJL}

\shorttitle{Delay of the r-process}
\shortauthors{Tarumi et al.}
\graphicspath{{./}{figures/}}

\begin{document}

\title{Evidence for r-process delay in very metal-poor stars}

\correspondingauthor{Yuta Tarumi}
\email{yuta.tarumi@phys.s.u-tokyo.ac.jp}

\author{Yuta Tarumi}
\affiliation{Department of Physics, School of Science, The University of Tokyo, Bunkyo, Tokyo 113-0033, Japan}

\author{Kenta Hotokezaka}
\affiliation{Research Center for the Early Universe, Graduate School of Science, University of Tokyo, Bunkyo-ku, Tokyo 113-0033, Japan}
\affiliation{Kavli IPMU (WPI), UTIAS, The University of Tokyo, Kashiwa, Chiba 277-8583, Japan}

\author{Paz Beniamini}
\affiliation{Division of Physics, Mathematics and Astronomy, California Institute of Technology, Pasadena, CA 91125, USA}
\affiliation{Astrophysics Research Center of the Open University (ARCO), The Open University of Israel, P.O Box 808, Ra’anana 43537, Israel}



\begin{abstract}
The abundances of $r$-process elements of very metal-poor stars capture the history of the $r$-process enrichment in the early stage of star formation in a galaxy. Currently, various types of astrophysical sites including neutron star mergers, magneto-rotational supernovae, and collapsars, are suggested as the origin of $r$-process elements.
The time delay between the star formation and the production of $r$-process elements is the key to distinguish these scenarios with the caveat that the diffusion of $r$-process elements in the interstellar medium may induce the delay in $r$-process enrichment because $r$-process events are rare.
Here we study the observed Ba abundance data of very metal-poor stars as the tracer of the early enrichment history of $r$-process elements. We find that the gradual increase of [Ba/Mg] with [Fe/H], which is remarkably similar among the Milky Way and classical dwarfs, requires a significant time delay
(100 Myr -- 1 Gyr)
of $r$-process events from star formation rather than the diffusion-induced delay. We stress that this conclusion is robust to the assumption regarding $s$-process contamination in the Ba abundances because the sources with no delay would overproduce Ba at very low metallicities even without the contribution from the {\it s}-process.
Therefore we conclude that sources with a delay, possibly neutron star mergers, are the origins of {\it r}-process elements.

\end{abstract}



\section{Introduction}
\label{sec:Intro}

Neutron-capture processes are responsible for the synthesis of the heaviest elements in the Universe \citep{2003Sneden}. 
\citet{1957_BBFH} classify them into two, namely the slow-process ({\it s}-process) and the rapid-process ({\it r}-process). The astrophysical sites that predominantly contribute to the production of {\it s}-process elements are the asymptotic giant branch (AGB) stars \citep{1957_BBFH, 1988Hollowell_TPAGB, 2004Travaglio_sprocess} while the origin of {\it r}-process elements remains one of the biggest  mysteries in nuclear astrophysics.

Recently, the observations of the gravitational wave signal and its electromagnetic counterpart from a neutron star merger (NSM), GW170817, provided us with evidence that NSMs eject copious amounts of $r$-process elements (see, e.g., \citealt{Margutti2020} for a recent review).
The merger rate and the mass of the ejected {\it r}-process elements derived from the observations of GW170817 agree remarkably well with those estimated from the $r$-process elemental abundances of Galactic stars  and  ultra-faint dwarf galaxies (UFDs) \citep{Eichler1989,Ji16_RetII,2016Beniamini_UFD,Cote2018, 2018Hotokezaka_Beniamini_rprocess_abundance,Rosswog2018} as well as geological measurements of radioactive isotopes \citep{hotokezaka2015,Tsujimoto2017,Bartos2019,Cote2020,2020Beniamini_diffusion}.

Even though the NSM scenario for the origin of $r$-process elements successfully explains the total amount of $r$-process elements in the Milky Way, it encounters potential obstacles to the delays between binary formation and merger. For example, a chemical evolution model with a delay time distribution of $\propto \Delta t^{-1}$ has difficulty explaining the distribution of stellar Eu abundances at higher metallicities ${\rm [Fe/H]}\gtrsim -1$ \citep{2018Hotokezaka_Beniamini_rprocess_abundance,2019Cote_onezone_MW}. 
To overcome this difficulty, several effects have been considered such as neutron star natal kicks, steep delay time distribution, and turbulent diffusion \citep{Wehmeyer2015MNRAS,Simonetti2019,Tsujimoto2019ApJ,Banerjee2020,Dvorkin2020}. This observation is one of the major arguments in support of $r$-process elements being predominantly produced by astrophysical phenomena other than NSMs, such as magneto-rotational supernovae (MRSNe,  \citealt{Nishimura15_MRSNe, Nishimura17_iprocess}), peculiar magnetar formation \citep{Metzger2008ApJ,Thompson2018}, common envelope jet supernovae \citep{2019NoamSoker}, and collapsars \citep{Siegel19_collapsar}. These latter phenomena are associated with the death of massive stars, and therefore they have no delay times between star formation and {\it r}-process enrichment.

A critical question for the origin of $r$-process elements is: do we have any direct observational evidence supporting the delay between star formation and the production of $r$-process elements, or are $r$-process elements preferentially produced in the early Universe? 
We can infer the history of metal enrichment in the Universe through the elemental abundance ratio and the overall metallicity of stars, e.g., the distribution of stellar abundances on a [{\it r}/Mg] - [Fe/H] plot.  
This paper aims to infer the time delay of {\it r}-process element production in a largely model-independent fashion and to clarify the origin of {\it r}-process elements. \citet{2004Ishimaru} assert that the origin of the {\it r}-process elements should not be the highest-mass stars using three stars with low Eu abundances at $\mathrm{[Fe/H]} \sim -3$. However, such measurements can be reconciled if $r$-process events are sufficiently rare because the  diffusion process delays the  chemical enrichment. 
Here we use Ba abundances of very metal-poor stars  ($\mathrm{[Fe/H]} < -2$) as the sample. 
With Ba abundances, which are available in a larger number of stars, we can distinguish the intrinsic delay of the $r$-process from the diffusion delay.
In section 2, we briefly describe the scenarios for the astrophysical phenomena suggested as the origin of $r$-process elements.  In section 3, we give the compilation of observed data and discuss the trend in the data. In section 4, we discuss the implication of the delay on the origin of the {\it r}-process in the Universe.

\section{Scenarios for the {\it r}-process and delay times}

    The scenarios for the origin of $r$-process elements are generally classified into three categories according to the relation between the star formation rate (SFR) and the production rate of $r$-process elements: (i) the delay scenario (NSM model), (ii) the no-delay scenario (Rare SN model), and (iii) the ``negative" delay scenario (Collapsar model; the origin of ``negative" delays is explained below). Any astrophysical scenario for the origin of $r$-process elements falls into one of the three categories. For all the scenarios we assume that $\alpha$-elements are predominantly produced by normal core-collapse SNe (ccSNe) and that the $r$-process events are rare, e.g., $\sim 1/1000$ of ccSNe. Here we briefly discuss these scenarios.
    
    {\it (i) The delay scenario (NSM model)}:
    A prominent feature of NSMs is a significant time delay after the formation of the progenitor stars. In fact, GW170817 occurred in a galaxy with weak star formation activity, suggesting that the time delay between the binary formation and merger is $\sim 1$--$10$ Gyr \citep{Blanchard2017,Levan2017}. The delay times of NSMs are widely distributed from a few tens of Myr to longer than a Hubble time, which is described by the delay-time distribution function, $DTD$:
    \begin{eqnarray}
    \dot{N}_r(t) \propto \int_0^t dt' DTD(\Delta t)SFR(t'),
    \end{eqnarray}
    where $\dot{N}_r$ is the rate of the $r$-process events, $\Delta t=t-t'$ is a time between binary neutron star (BNS) formation and merger, and $SFR$ is the star formation rate.
    The form of $DTD$ is often 
    assumed to be $\propto \Delta t^{-1}$ with a minimum delay $\Delta t_{\rm min}$. \cite{2019Beniamini_FastMergingBNS} find a somewhat steeper power law, $DTD\propto  \Delta t^{-1.3}$ with $\Delta t_{\rm min} \sim 35\,{\rm Myr}$, inferred from the orbital separations of the Galactic BNS. In addition, the redshift distribution of short gamma-ray bursts (GRBs) is consistent with a ${ DTD} \propto \Delta t^{-1}$ with $\Delta t_{\rm min}\approx 20\,{\rm Myr}$ \citep{Wanderman2015}.

   {\it (ii) The no-delay scenario (Rare SN model)}: 
   Peculiar ccSNe such as MRSNe and peculiar magnetar formation are proposed to produce heavy $r$-process elements \citep{Nishimura15_MRSNe,Metzger2008ApJ,Thompson2018}. 
   Since the evolution of the fraction of these peculiar ccSNe to normal ccSNe evolves is unknown, we assume here that it is constant with time during the Galaxy's evolution. The production rate of $r$-process elements is then proportional to that of $\alpha$-elements, i.e., $\dot{N}_r(t)\propto SFR(t)$.

   {\it (iii)  The ``negative" delay scenario (Collapsar model)}: \cite{Siegel19_collapsar} propose that the massive outflow from the central engine of long GRBs, i.e., collapsars, can be  the site of the $r$-process\footnote{The nucleosynthesis of heavy elements in collapsars is still under debate \citep{Fujibayashi2020}.}. A ``negative" delay is somewhat counter-intuitive, when discussing $r$-process enrichment associated with the death of massive stars (which of course occurs after their birth, not before). It should instead be understood in an averaged sense. ``Negative" delays from star formation to $r$-process enrichment are a result of the fact that long GRBs preferentially occur in environments with large specific SFR and low metallicities \citep{Svensson2010MNRAS,Palmerio2019A&A}. This is perhaps most clearly demonstrated by the redshift distribution of long GRB which peaks at a  higher redshift than the cosmic SFR \citep{Wanderman}. 
   Therefore, we assume that the production ratio of $r$-process elements to $\alpha$-elements decreases with time in this scenario:
   \begin{eqnarray}
    \dot{N}_r(t) \propto A(t)SFR(t),
   \end{eqnarray}
   where $A(t)$ is a decreasing function, which takes into account the enhancement of the event rate of collapsars at the earlier times. In what follows, we assume $A(t)\propto t^{-0.5}$ motivated by the relation between the long GRB rate and cosmic SFR at $z\lesssim 3$ \citep{Wanderman}. We caveat that the rate of long GRB in the very early Universe is not well constrained. We note, however, that the ``negative" delay scenario is not necessarily limited to collapsars and would apply to other sources which peak at greater $z$ than the Cosmic star formation.
    
\section{{\it r}-process delay inferred from very metal-poor stars}    

    \begin{figure}
    \centering
    \includegraphics[width=\columnwidth]{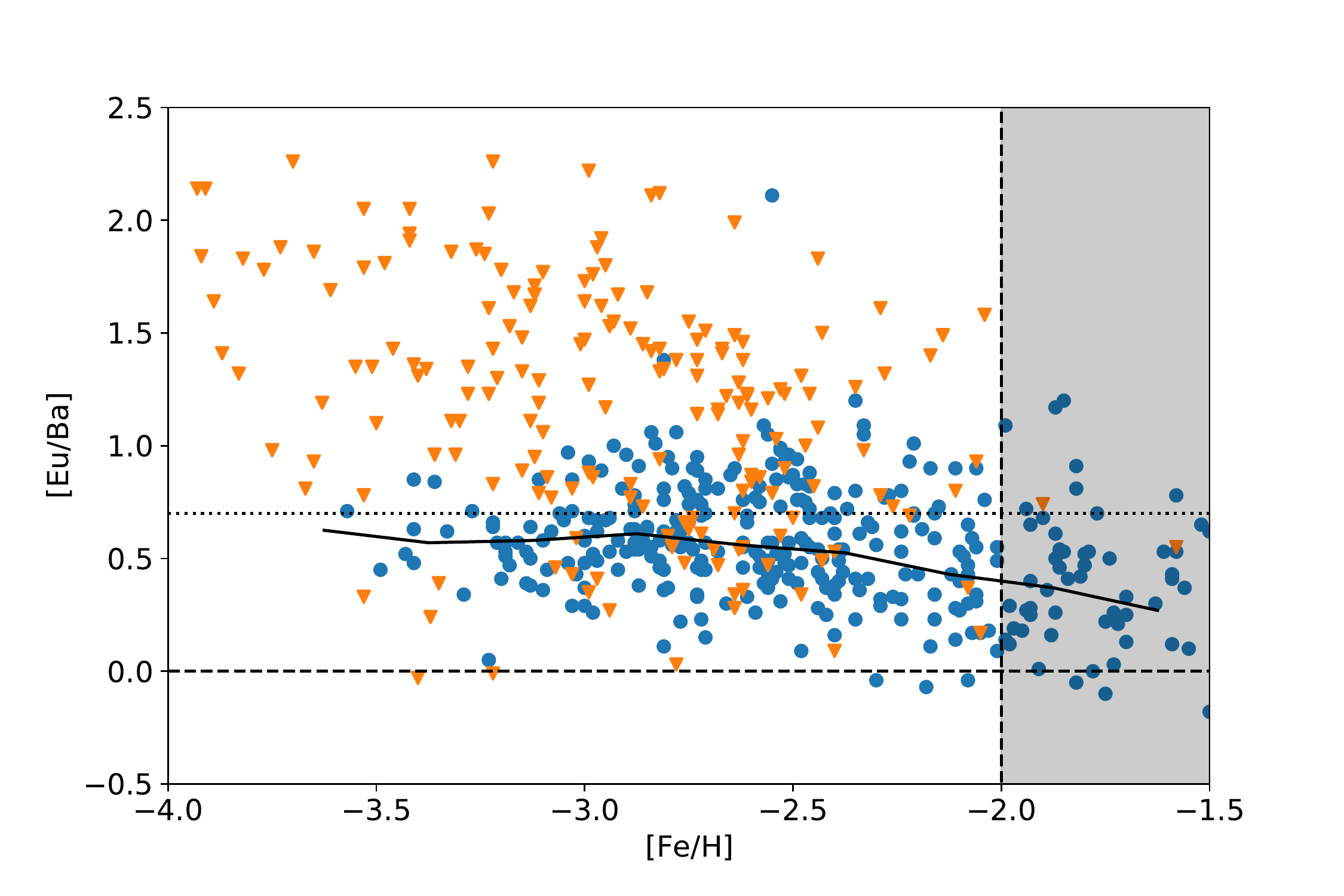}
    \caption{[Eu/Ba] evolution at the lowest metallicity (upper limits in orange triangles, measurements in blue points). The dotted line shows the {\it r}-process value ([Eu/Ba] = 0.7; \citealt{2008Sneden_review}). Stars with $[\mathrm{C}/\mathrm{Fe}] < 0.5$ are plotted. For $[\mathrm{Fe}/\mathrm{H}] < -2$ stars, the evolution of [Eu/Ba] is only $\sim 0.2$ dex, meaning that the contribution of the {\it s}-process to the Ba production is not dominant.}
    \label{fig:EuBa_FeH}
    \end{figure}

    We now turn to constrain the early enrichment history of $r$-process elements by using the {\it r}-process abundances of very metal-poor stars.
    Europium (Eu) is the most commonly used as the {\it r}-process tracer element. However, since the absorption lines of Eu are not so strong, the Eu measurements are often unavailable for very metal-poor stars. In such cases, barium (Ba) abundances are used as the tracer of $r$-process elements instead of Eu \citep{fran2007A&A,Duggan2018ApJ}. Here we study Ba abundance evolution at the low metallicity range of $-4\lesssim {\rm [Fe/H]}\lesssim -2$. Although Ba is also produced by the {\it s}-process, the production is dominated by the {\it r}-process at this metallicity range. Figure~\ref{fig:EuBa_FeH} shows the [Eu/Ba] of metal-poor stars in the Milky Way. The value of [Eu/Ba] gradually decreases as the contribution of {\it s}-process sets in at $[\mathrm{Fe}/\mathrm{H}] \gtrsim -2.0$. Stars with metallicity lower than that have values of [Eu/Ba] consistent with the {\it r}-process. 
    \footnote{The exception is the {\it s}-enhanced carbon-enhanced metal-poor (CEMP-{\it s}) stars. These stars are part of binaries, and the compositions of the stars are dominated by mass-transfer from the companions. Such stars are not included in this work.}. 

\begin{figure}
    \centering
    \includegraphics[width=\columnwidth]{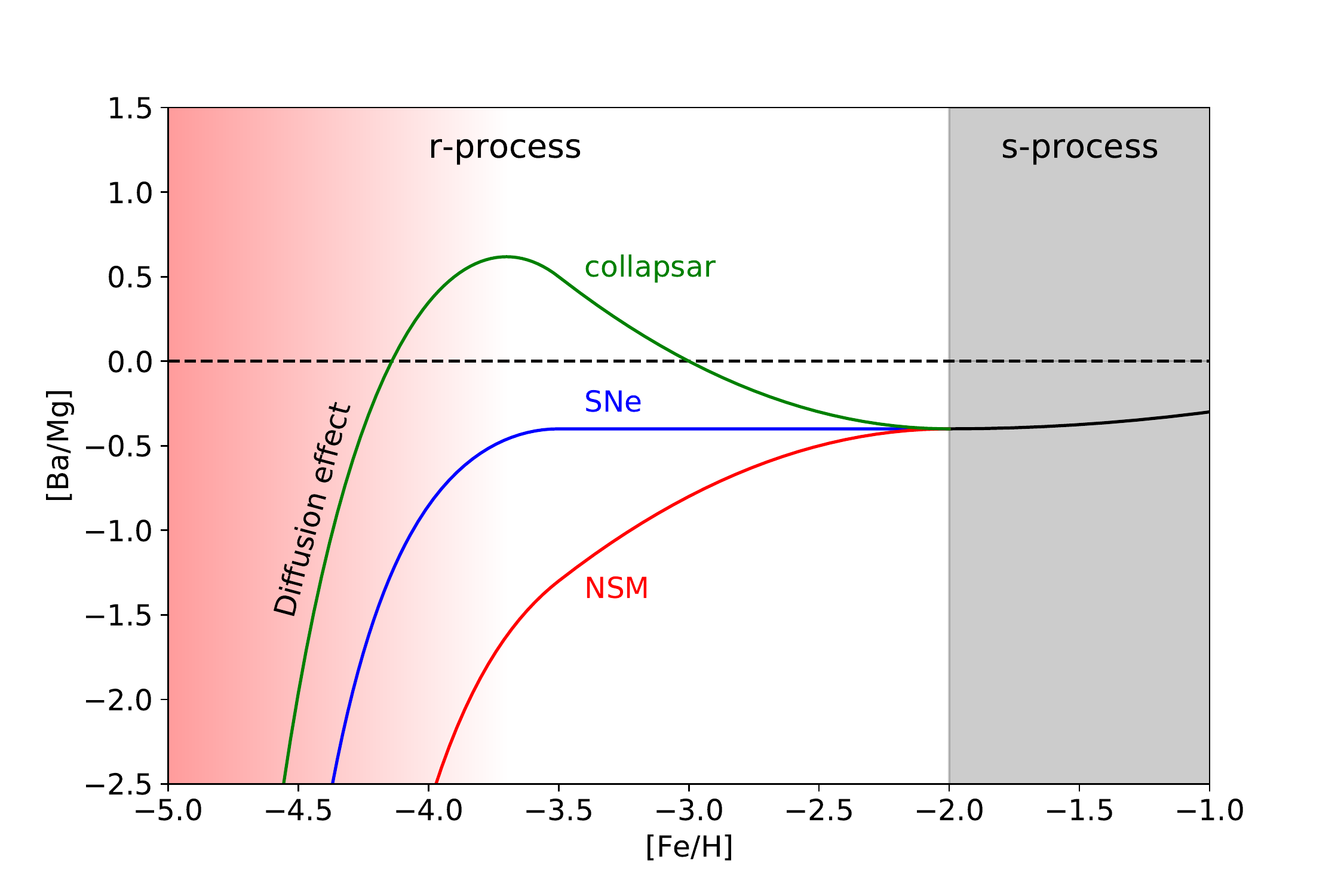}
    \caption{
    A schematic figure showing the evolution of [Ba/Mg] against [Fe/H]. A sharp increase of the median [Ba/Mg] is expected at very low [Fe/H]. This is as a result of the inhomogeneity effect described in the text, which holds for times short relative to the mixing time, where the Ba enrichment at a given location is largely dominated by one event. At $-3.5 \lesssim \mathrm{[Fe/H]} \lesssim -2.0$ the inhomogeneity effect is weak and the [Ba/Mg] trend is determined by the delay of {\it r}-process enrichment. At $-2.0 \lesssim \mathrm{[Fe/H]}$ contributions from {\it s}-process becomes important and we see the increase in [Ba/Mg] with increasing [Fe/H].}
    \label{fig:Ba evolution}
\end{figure}
    
    Figure~\ref{fig:Ba evolution} shows a schematic figure of the [Ba/Mg] evolution against [Fe/H]. The Galaxy was metal-poor at early times, and as a result, stars formed very early shape the curve on the left. At such extremely low metallicities, the chemical inhomogeneity of the interstellar medium (ISM) plays an important role because the event rate of the $r$-process is much lower than normal SNe. The {\it r}-process elements in the ISM at a given location originate from a single enrichment event \citep{2020Beniamini_diffusion}. The volume fraction of the highly enriched ISM is very small and therefore typical stars (outside of the enriched areas) have very low [Ba/Mg]. This inhomogeneity effect quickly disappears  as more events start to contribute to the enrichment at each location, and the [Ba/Mg] track now represents the intrinsic delay of {\it r}-process element production. At $\mathrm{[Fe/H]} \gtrsim -2$, the {\it s}-process starts to contribute significantly to the Ba abundances. As we are interested in constraining the $r$-process enrichment, in what follows, we will focus on very metal poor stars with $\mathrm{[Fe/H]} \lesssim -2$.
    
\begin{figure*}
    \centering
    \includegraphics[width=2\columnwidth]{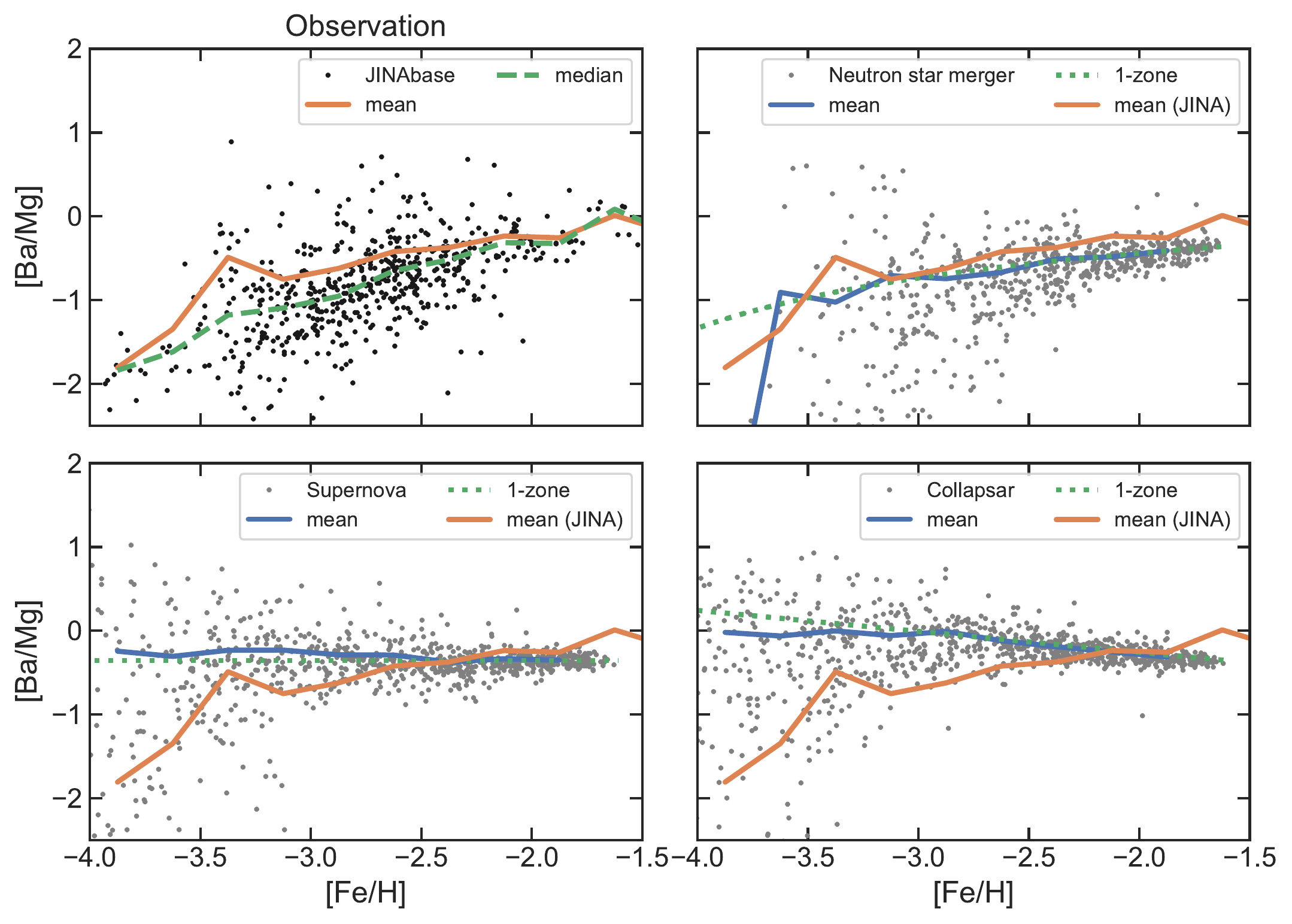}
    \caption{
    Evolution of [Ba/Mg] against [Fe/H] in observed stars and in our diffusion model. Top-left panel: observed distribution. The data is obtained in JINAbase \citep{2018JINA}. Stars with high carbon abundance ($\mathrm{[C/Fe]} > 0.5$) are excluded. 
    The three other panels are the synthetic observations of our model. Top-right panel: results of the NSM model. Bottom-left panel: results of SNe model. Bottom-right panel: results of the collapsar model. The SFR is $\propto t^{\alpha}$ and here we choose $\alpha=1$ in all the cases. For the NSM model, we use a power-law DTD $\propto t^{-1}$ with a minimum delay time of $20 \mathrm{Myr}$. For the SNe model {\it r}-process enrichment occurs concurrently with star formation. For the collapsar model we assume the rate to be proportional to $SFR(t)\times t^{-0.5}$, so that the {\it r}-process enrichment preferentially occur before the Fe and Mg enrichment. The diffusion coefficient $D$ is set to $D = 0.2\ \mathrm{kpc^{2}/Gyr}$ for all the models. The curves labeled by $1$-zone are the result of the one-zone modeling, in which the instantaneous mixing is assumed, i.e., $D\rightarrow \infty$.}
    \label{fig:Model comparison}
    \end{figure*}

    The top-left panel of Figure~\ref{fig:Model comparison} plots the abundances of the observed metal poor stars.
    To quantitatively discuss the implications, we over-plot two curves, ``linear mean'' and median. The linear mean $\mu(\mathrm{Ba}/\mathrm{Mg})$ is defined as:
    \begin{equation}
        \mu(\mathrm{Ba}/\mathrm{Mg}) =\frac{1}{N_\mathrm{star}} \sum_{i}^{N_\mathrm{star}} 10^{[\mathrm{Ba}/\mathrm{Mg}]_{i}},\label{eq:mean}
    \end{equation}
    where the index $i$ runs from 1 to the number of stars in each [Fe/H] bin. 
    We ignore stars with upper limits. The curves were mostly unchanged even if we assume all upper limit stars to have $\mathrm{[Ba/Mg]} = -10$.

    The interpretation of the increasing trend of [Ba/Mg] with [Fe/H] has to be treated with caution. Although the time delay would certainly be the reason for the increase, the inhomogeneity effect could also make a pseudo-increasing trend on the median even without time delay, as we have shown in Figure~\ref{fig:Ba evolution}. This is because the abundance distribution has a long tail on the high-abundance end. The skewed abundance distribution naturally arises in the elemental abundance distribution of the ISM in a galaxy. Initially, one {\it r}-process event enriches a small portion of the ISM to a very high degree, and it dilutes to the whole galaxy as time passes. If we measure the distribution of the {\it r}-process element abundances of the ISM before complete mixing, the median would always be less than the ``one-zone'' prediction, for which instantaneous mixing is assumed.
    
    In contrast, the linear mean, $\mu([\mathrm{Ba}/\mathrm{Mg}])$, is robust to the inhomogeneity of the distribution of {\it r}-process elements. 
    It gives appropriate weights to both {\it r}-rich and {\it r}-poor gas. The deviation from a one-zone model arises because we are comparing the medians, which overestimate the importance of volumetrically dominant {\it r}-poor gas.
    If we assume that Mg and Fe distributions are homogeneous, and we sample the whole system with a sufficient resolution, the linear mean exactly matches the one-zone model with the instant mixing approximation. Therefore we conclude that the increase in the linear mean $\mu([\mathrm{Ba}/\mathrm{Mg}])$ is a result of the time delay between the production of Mg and $r$-process elements. The remaining problems are the validity of the assumptions. Namely, (i) the homogeneity of the denominator (Mg) and the x-axis (Fe), and (ii) a sufficient sampling of the volume. Although validating these conditions is difficult, we expect that such effects would deviate the linear mean symmetrically from the one-zone calculation. Therefore it is unlikely that the inhomogeneity effect alone gives the clear increasing trend suggested by the observations.
    
    An independent argument is the gradual increase of the median [Ba/Mg]. As we explained above, the inhomogeneity effect can make an increasing trend in the median. However, the increase occurs exponentially, and therefore it would be very abrupt in this case.
    We can estimate the typical [Fe/H] range of the increase by the total volume fraction of {\it r}-rich bubbles. Here we estimate the range ($\Delta[\mathrm{Fe}/\mathrm{H}]$) from $[\mathrm{Ba}/\mathrm{Mg}] \sim -2.0$ to $[\mathrm{Ba}/\mathrm{Mg}] \sim -0.5$ assuming that the one-zone value [Ba/Mg]$_\mathrm{1zone}$ is $\sim -0.5$. The volume of each {\it r}-rich bubble ($1 \sigma$ region) is $\sim \sigma^{3}$. The median of [Ba/Mg] roughly converges to [Ba/Mg]$_\mathrm{1zone}$ when the total volume fraction of {\it r}-rich bubbles becomes $\sim 1$. 
    The typical volume of each bubble and the number of bubbles grow as $\propto t^{3/2}$ and $t^{\alpha+1}$, respectively. Here we have assumed $(SFR\ density) \propto t^{\alpha}$. Thus the total volume fraction $V_\mathrm{tot, 1 \sigma}$ grows as $\propto t^{\alpha+5/2}$. Since [Fe/H] increases logarithmically with the number density of stars formed, we have $[\mathrm{Fe}/\mathrm{H}] \sim (1+\alpha)\log_{10}(t)+const.$ Therefore, the value of [Fe/H] when the [Ba/Mg] converges to [Ba/Mg]$_\mathrm{1zone}$ is
    \begin{equation}
        [\mathrm{Fe}/\mathrm{H}] \approx \log_{10}(V_\mathrm{tot, 1 \sigma})\cdot \frac{(\alpha+1)}{(\alpha+5/2)}+const.
    \end{equation}
    We can follow the same procedure to derive the [Fe/H] when the median of [Ba/Mg] surpasses [Ba/Mg]$_\mathrm{1zone} - 1.5$ by substituting $V_\mathrm{tot, 3 \sigma} = 3^{3} \cdot V_\mathrm{tot, 1 \sigma}$ for $V_\mathrm{tot, 1 \sigma}$\footnote{Assuming a Gaussian distribution, $10^{-1.5}\times e^{-0.5} \simeq e^{-3.95} = e^{-7.9/2}$, therefore a region of size $\sqrt{7.9} \sigma \simeq 3 \sigma$ has [Ba/Mg] value higher than [Ba/Mg]$_\mathrm{1zone} - 1.5$. The volume of a $3 \sigma$ region is $3^3$ times that of the $1 \sigma$ region. Therefore the total volume fraction of the $3 \sigma$ regions becomes 1 when that of $1 \sigma$ regions is $1/3^{3}$.}. Finally, $\Delta \mathrm{[Fe/H]}$ can be derived as 
    $\Delta[\mathrm{Fe}/\mathrm{H}] \approx \log_{10}(3^{3})\cdot (\alpha+1)/(\alpha+5/2)$, and thus, we obtain the range that the diffusion increase occurs as $0.56\lesssim \Delta[\mathrm{Fe}/\mathrm{H}] < 1.4$ for  $\alpha>0$.
    However, the observational data show a slower increase, 
    $\Delta [\mathrm{Fe}/\mathrm{H}] \sim 2.0$,
    which a diffusion-induced increase cannot reproduce.

    To demonstrate the effects we have discussed, we run a Monte Carlo simulation of the chemical enrichment of the early phase of the Milky Way evolution, $<0.5\,{\rm Gyr}$. Note that our model is also applicable to the chemical evolution of other galaxies simply by adopting another SFR evolution. Although our model does not capture outflow from galaxies \citep{2018Emerick}, we expect this not to be problematic as we are discussing short timescales. Figure~\ref{fig:Model comparison} contains three panels from our model calculations. Here we briefly summarize our synthetic models. We prepare a ${\rm kpc^3}$ box filled with hydrogen with a constant density, $\approx 20\,{\rm cm^{-3}}$. The SFR is taken as $SFR \propto t^{\alpha}$, and is normalized such that the mean ccSN rate is $10^{3}\,{\rm Myr^{-1}kpc^{-3}}$.  Fe and Mg are injected randomly in space with a rate proportional to the SFR.
    We assume that each ccSN produces $m_{\rm Fe}=0.04M_{\odot}$ (e.g. \citealt{Nakar2016ApJ}) and $m_{\rm Mg}=0.05M_{\odot}$ and each r-process event produces $m_{\rm Ba}=3\cdot 10^{-4}M_{\odot}$.
    We fix the overall rate of the {\it r}-process events at $\sim 1/1000$ of ccSNe \citep{2016Beniamini_UFD,2018Hotokezaka_Beniamini_rprocess_abundance}.

    We follow the dilution of the produced Fe, Mg, and {\it r}-process elements in the ISM via turbulent mixing with the method used in \citet{hotokezaka2015,2020Beniamini_diffusion}. The number density of an element $X$ at position $\vec{r}$ at time $t$ is:
    \begin{equation}
        n_X(\vec{r}, t) = \sum_{t_{j}<t-t_{*}} \frac{N_{X}}{K_{j}(\Delta t_{j})}\exp\biggl[-\frac{|\vec{r}-\vec{r}_{j}|^{2}}{4D\Delta t_{j}} \biggr]
    \end{equation}
    where $N_X$ is the total number of the element $X$ produced by each event,
    $t_{j}, \vec{r}_{j}$ are time and spatial position of the $j$-th event, $\Delta t_{j} = t - t_{j}$, $K_{j}(\Delta t_{j}) = (4\pi D\Delta t_{j})^{3/2}$ and $D$ is the diffusion coefficient\footnote{Once $K_j$ reaches the volume of the box, we fix $K_j$ in order to avoid an artificial leakage of the elements \citep{2020Beniamini_diffusion}. }. 
    Following the results of \cite{2020Beniamini_diffusion} we set the $D$ to $0.2\ \mathrm{kpc^{2}/Gyr}$ for the simulation. We track stars formed at random positions in the box. Their abundances are given by the ISM abundances at their birthplaces and times of birth.
    
    Figure~\ref{fig:Model comparison} shows the results of synthetic data obtained with our models. Also shown is the abundance evolution obtained by using the one-zone models and the linear mean  calculated with the same procedures as the real data (equation \ref{eq:mean}). 
    As we stated above, the mean traces the one-zone prediction in all the models.  The SN and Collapsar models overproduce stars with ${\rm [Ba/Mg]>0}$ compared with the observed distribution. 
    It is clear that this overproduction of Ba cannot be solved by adding the $s$-process contribution.
    Furthermore, it is also clear from the figure that
    the [Ba/Mg] very quickly converges to the one-zone lines within $\sim 0.5\,{\rm dex}$ in [Fe/H]. Therefore
    the observed gradual rise in [Ba/Mg] results from the intrinsic time delay in the production of $r$-process elements relative to the production of Mg, i.e., $r$-process events' delay from ccSNe. 
    This result supports the idea that NSMs are the source of 
    $r$-process elements in the low metallicity region ${\rm [Fe/H]}\lesssim -2$. We also performed a simulation with $DTD\propto \Delta t^{-1.3}$ motivated by \cite{2019Beniamini_FastMergingBNS} and obtain a result similar to that with $DTD\propto \Delta t^{-1}$.
    Finally, we also examine [Eu/Mg] in figure \ref{fig:EuMg}. While the stars with Eu detection do not show a clear increasing trend in [Eu/Mg] compared to [Ba/Mg]. However, this is consistent with the rise in [Ba/Mg] if the stars with upper limits on Eu are accounted for.   
    
\begin{figure}
        \centering
        \includegraphics[width=\columnwidth]{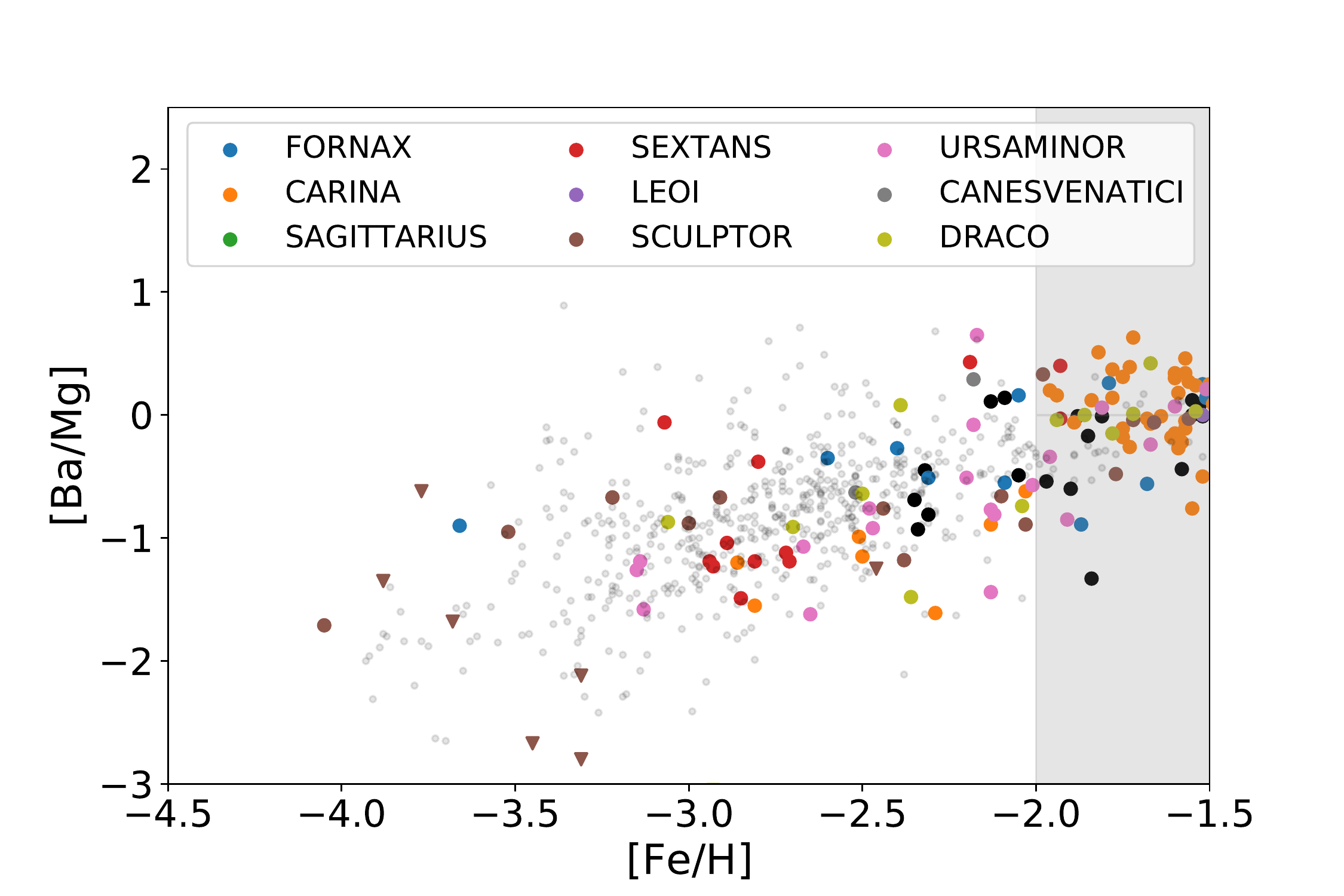}
        \caption{[Ba/Mg] - [Fe/H] plot for classical dwarf galaxies. Circles are detections and downward triangles are upper limits. The data for these galaxies are obtained from the SAGA database \citep{SAGA_dwarf}. 
        The gray dots are stars in the MW, and the black dots are mean abundances of globular clusters in MW halo \citep{2005Pritzl_MWandSgr_GC}.
        The dwarf galaxies are on the same track of the Milky Way stars, supporting the interpretation that the gradual rise of [Ba/Mg] is caused by the intrinsic delay of $r$-process events.
        }
        \label{fig:dwarf}
    \end{figure}

    Figure~\ref{fig:dwarf} shows the [Ba/Mg] distribution for stars in classical dwarf galaxies.  Remarkably, the same increasing [Ba/Mg] - [Fe/H] relation holds also for dwarf galaxies despite the fact that the time scale of chemical enrichment of the dwarf galaxies is longer than that of the MW, although the number of data points may not be very compelling. One can expect that such a similarity arises if (i) the time dependence of their SFHs, i.e. the values of $\alpha$, are similar, (ii)  the DTD is very approximately $\propto \Delta t^{-1}$,  (iii) $t_{\rm min}$ is shorter than the time scale of chemical enrichment  of the galaxies, e.g., the time scale on which a galaxy evolves to ${\rm [Fe/H]}\sim -3$, (iv) the $s$-process and SNe Ia have not contributed significantly to the Ba and Fe abundances, respectively. Therefore the similarity seen in Fig. \ref{fig:dwarf} supports our conclusion of  $r$-process delay and suggests that the scenarios with a constant delay time such as $\Delta t = 100\,{\rm Myr}$ are less likely. We have ignored here the outflow efficiency that may be different between galaxies and have a significant impact on the stellar abundances of a galaxy \citet{2018Emerick}. Even without considering this effect, the similarity persists robustly. 
    Note that here we omit UFDs because they are so small that  they likely have experienced none or only one $r$-process event in the past \citep{2016Beniamini_UFD}. In this case, the distribution of the stellar $r$-process abundances is significantly different from that of galaxies that experienced multiple events and does not reflect the delay
    \citep{Safarzadeh2017MNRAS,Beniamini_retain_2018,Tarumi2020MNRAS}.

\section{Conclusions and discussions}

    The observed abundance distribution of [Ba/Mg] of the Galactic very metal poor stars clearly shows that [Ba/Mg] significantly increases as [Fe/H] increases at $\mathrm{[Fe/H]} < -2.0$. 
    Such an increase is attributed to the intrinsic time delay of the $r$-process events and/or the delay induced by chemical diffusion in the ISM.
    We showed that the increase of the linear mean and the gradual increase of the median  require an intrinsic time delay of the production of $r$-process elements and cannot be explained solely by the diffusion-induced delay. Therefore we conclude that {\it r}-process element production occurs with significant delays at this metallicity range, which is naturally expected in the NSM scenario (Figure \ref{fig:Model comparison} and  see also \citealt{Matteucci2014MNRAS,Hirai2015ApJ,Wehmeyer2015MNRAS,2019Cote_onezone_MW,Simonetti2019,Voort2020MNRAS}).
    Any rare SN such as MRSNe and collapsars are not expected to have a significant time delay between the {\it r}-process production and Mg production. 
    Consequently, SN models overproduce Ba at very low metallicities even if the contribution of the $s$-process to the Ba abundances is completely ignored. 
    Therefore such sources are incompatible with the observed trend.
    We conclude that this is a new line of evidence supporting  NSMs as the origin of {\it r}-process elements at least in low metallicity environments [Fe/H] $\lesssim -2$. Note that it is important to use Ba as the tracer. If we do the same analysis for Eu only with the detected samples, the increasing trend we have discussed is buried under noisy behavior (see \S  \ref{sec:EuMg}). By including upper limits as zero abundance, we see a trace of the increasing trend as in Ba.

    The [Ba/Mg] abundances of very metal poor stars in some dwarf galaxies also support the existence of significant time delays. Although the sample is small and dwarf galaxies have a different star formation history, most stars are on the same track as the Milky Way stars  (Figure \ref{fig:dwarf} and see also \citealt{Reichert2020A&A}). Therefore the ``delay'' is likely a common feature, which can be tested with  larger samples of the Ba abundances of dwarf galaxies. 
    
    Note that Ba is detected in most metal-poor stars as pointed out by \citet{2013Roederer_BaSrFloor}. The apparent ``floor'' of Ba abundance might indicate that some unknown neutron-capture processes are working in addition to the {\it r}-process source. The existence of such an event is also supported from Ba abundances of UFDs \citep{2020Tarumi_sprocess}.
    In this case, the intrinsic increase of [{\it r}/Mg] may be steeper than suggested by the Ba observations. However, our argument regarding the increasing mean still holds: stars with such low Ba abundances do not affect the linear mean [Ba/Mg] anyway.
    
    It is possible that a significant fraction of the very metal-poor stars originally formed in dwarf galaxies that are now disrupted. Even in this case our argument remains the same. As Figure~\ref{fig:dwarf} shows, the secular evolution is present in each dwarf galaxy. To explain the ubiquitous increasing trend among dwarf galaxies, a delay in {\it r}-process enrichment is necessary.
    
    Some globular clusters (GCs) show internal spread of {\it r}-process element abundances. The short duration of the star formation makes it difficult for delayed sources to occur within active star formation. However, considering the small internal spread of metallicity within each GC, the difficulty is common also to other origins. It is natural to expect that the star formation in a GC quenches before massive stars contribute to its chemical enrichment. Therefore no {\it r}-process events can contribute to the enrichment of this system. One possible solution, is the Eu spread in ISM. The formation process of GCs is poorly understood, and an investigation of the origin of such {\it r}-process spread would be interesting.

    Further investigation would be needed to explain the $[\mathrm{Eu}/\mathrm{Mg}] \sim 0$ trends seen at the higher metallicities (e.g. \citealt{2020_GALAH_DR3} for stars in the Milky Way), which supports shorter or no time delay of the {\it r}-process events. One difference from the lower [Fe/H] regime is the timescale. At the relatively higher metallicities, [$\mathrm{Fe}/\mathrm{H}$] $\gtrsim -1$, where $[\mathrm{Eu}/\mathrm{Mg}] \sim 0$ is seen, it reflects delays of $1$--$10\,\mathrm{Gyr}$. A similar feature is also found for the Sculptor dwarf galaxy.
    At $-2 < [\mathrm{Fe}/\mathrm{H}] < -1$ where  \cite{2019Skuladottir_Sculptor} report $[\mathrm{Eu}/\mathrm{Mg}] \sim 0$, they also found increasing $[\mathrm{Ba}/\mathrm{Mg}]$, which is clear evidence of the delayed contribution from AGB stars. 
    Recently,  \cite{Matsuno2021} studied the $r$-process enrichment of Gaia-Enceladus stars. While they found a significant excess in the $r$-process abundance normalized by an $\alpha$-element, i.e., $[\mathrm{Eu}/\mathrm{Mg}] \sim 0.3$, $[\mathrm{Eu}/\mathrm{Mg}]$ does not evolve with [Fe/H].
    The fact that {\it r}-process delays are seen at $[\mathrm{Fe}/\mathrm{H}] \lesssim -2$ and not at $[\mathrm{Fe}/\mathrm{H}] \gtrsim -1$ may indicate that the typical time delay of NSM is the order of a few $ 100\ \mathrm{Myr}$, consistent with constraints from the Galactic BNS population \citep{2019Beniamini_FastMergingBNS}. 
    
    The typical delay time of NSM is comparable to or longer than the one of AGB stars. The reason that significant contributions of AGB stars come in only at higher metallicity is likely the metallicity dependence of the {\it s}-process. At the lowest metallicity of $[\mathrm{Fe}/\mathrm{H}] \lesssim -2$, the amount of seed nuclei for the {\it s}-process nucleosynthesis is small (see, e.g. \citealt{2011Kappeler_sprocess_review}). Therefore, it is naturally expected and consistent with observation that the contribution of AGB stars comes in later than the {\it r}-process in NSM.
    
    As mentioned in \S \ref{sec:Intro}, it was recently suggested that the natal kicks of BNS can play an important role in shaping the $r$-process abundance distribution \citep{Banerjee2020}. This requires the typical center of mass kicks to be $\Delta v_{\rm cm}\gtrsim 100\ \mbox{km s}^{-1}$. This, however, is in contrast with an analysis of the kinematics of Galactic BNS which suggests that most systems were born with $\Delta v_{\rm cm}<10\ \mbox{km s}^{-1}$ and finding the inferred kicks to be uncorrelated with the observed GW merger times \citep{BP2016}. Indeed if kicks had played an important role in the MW chemical evolution, it would have been more difficult to explain the observed $r$-process abundance in Reticulum II (see \citealt{2016Beniamini_natalkicks}), whereas the measured abundances in Reticulum II and in classical dwarfs are fully in agreement with estimates from BNS \citep{2016Beniamini_UFD}. An independent constraint comes from short GRBs. The offset distribution of those GRBs relative to the centers of their host galaxies is found to be fully consistent with the underlying distribution of stellar material in the host galaxies \citep{PB2021}. This strongly suggests that the effects of kicks are subdominant in determining the merger locations of BNS.
    
    In summary, we found that a significant increase of [{\it r}/Mg] at the lowest metallicity stars requires a delay of $100\ \mathrm{Myr} \sim 1\ \mathrm{Gyr}$  between star formation and the production of $r$-process elements. No-delay and negative delay sources are incompatible with the observed data, even if we consider the inhomogeneity effect. Since NSMs naturally introduce a comparable delay, it is promising as the origin of {\it r}-process in the Universe.

\acknowledgments

We thank the anonymous referee for useful comments.
We also thank Yutaka Hirai, Tadafumi Matsuno, Naoki Yoshida, Brian Metzger, and Daniel Siegel for insightful comments.
Y. T. is supported by JSPS KAKENHI Grant Number 20J21795. K. H. is supported by JSPS Early-Career Scientists Grant Number 20K14513. The research of P. B. was funded by the Gordon and Betty Moore Foundation through Grant GBMF5076.

%

\vspace{5mm}





\appendix

\section{Dependence on star formation history}
\begin{figure}
    \centering
    \includegraphics[width=0.5\columnwidth]{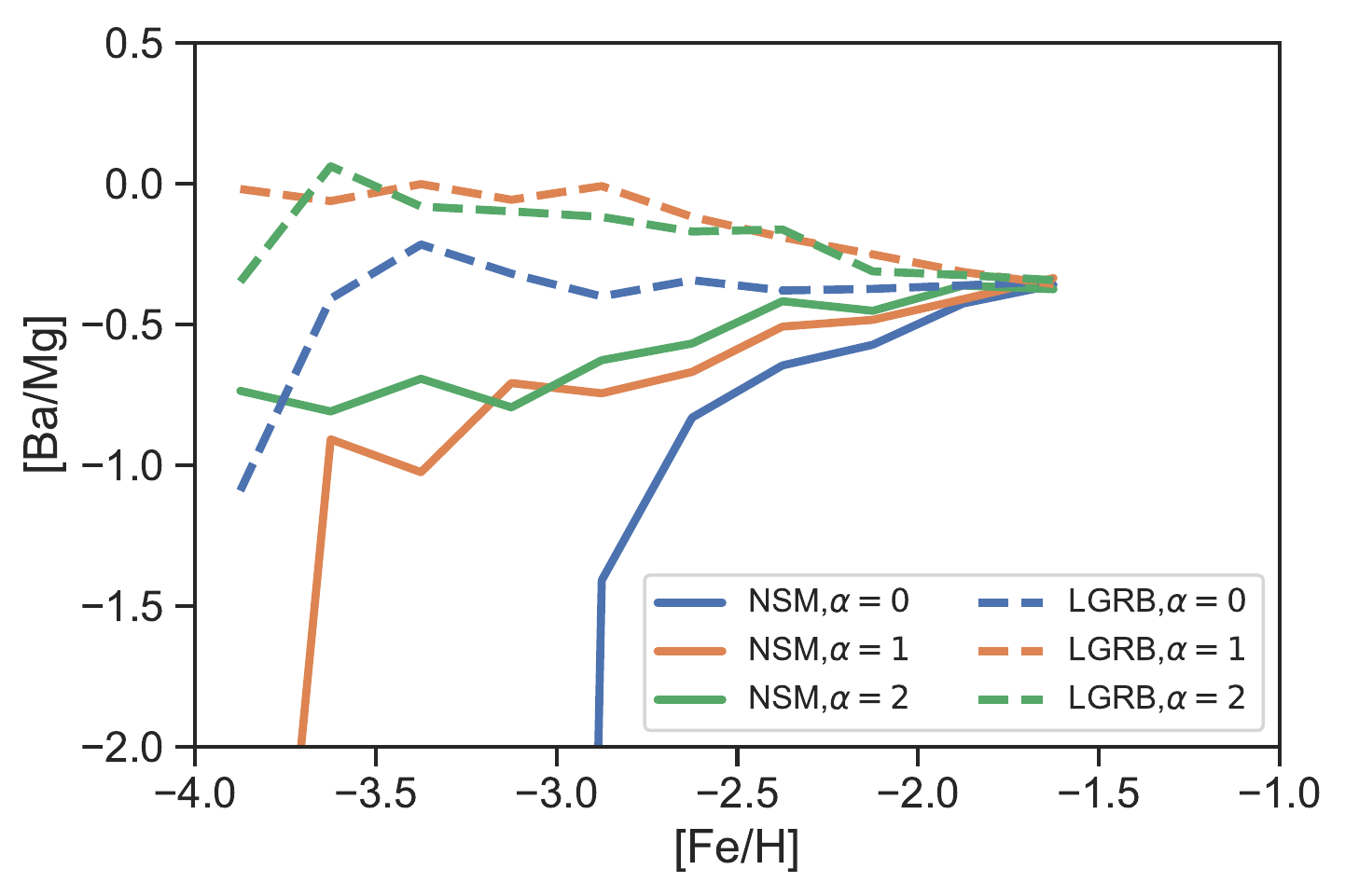}
    \caption{The evolution of [Ba/Mg] values in various star formation histories. The diffence of star formation history is color-coded. Solid lines and dashed lines show results with the NSM model and the collapsar model.}
    \label{fig:SFH}
\end{figure}

In Figure~\ref{fig:SFH} we show the evolution of mean [Ba/Mg] with three different star formation histories. Solid lines and dashed lines show results with the NSM model and the collapsar model. The NSM models robustly show the increasing trends whereas the collapsar models show decreasing trends. Considering the observed similarity of the increasing trends between the MW and dwarf galaxies (Figure~\ref{fig:dwarf}), NSM is favored as the origin of the {\it r}-process for either of those environments.

\section{Analysis with Eu}
\label{sec:EuMg}

\begin{figure}
    \centering
    \includegraphics[width=\columnwidth]{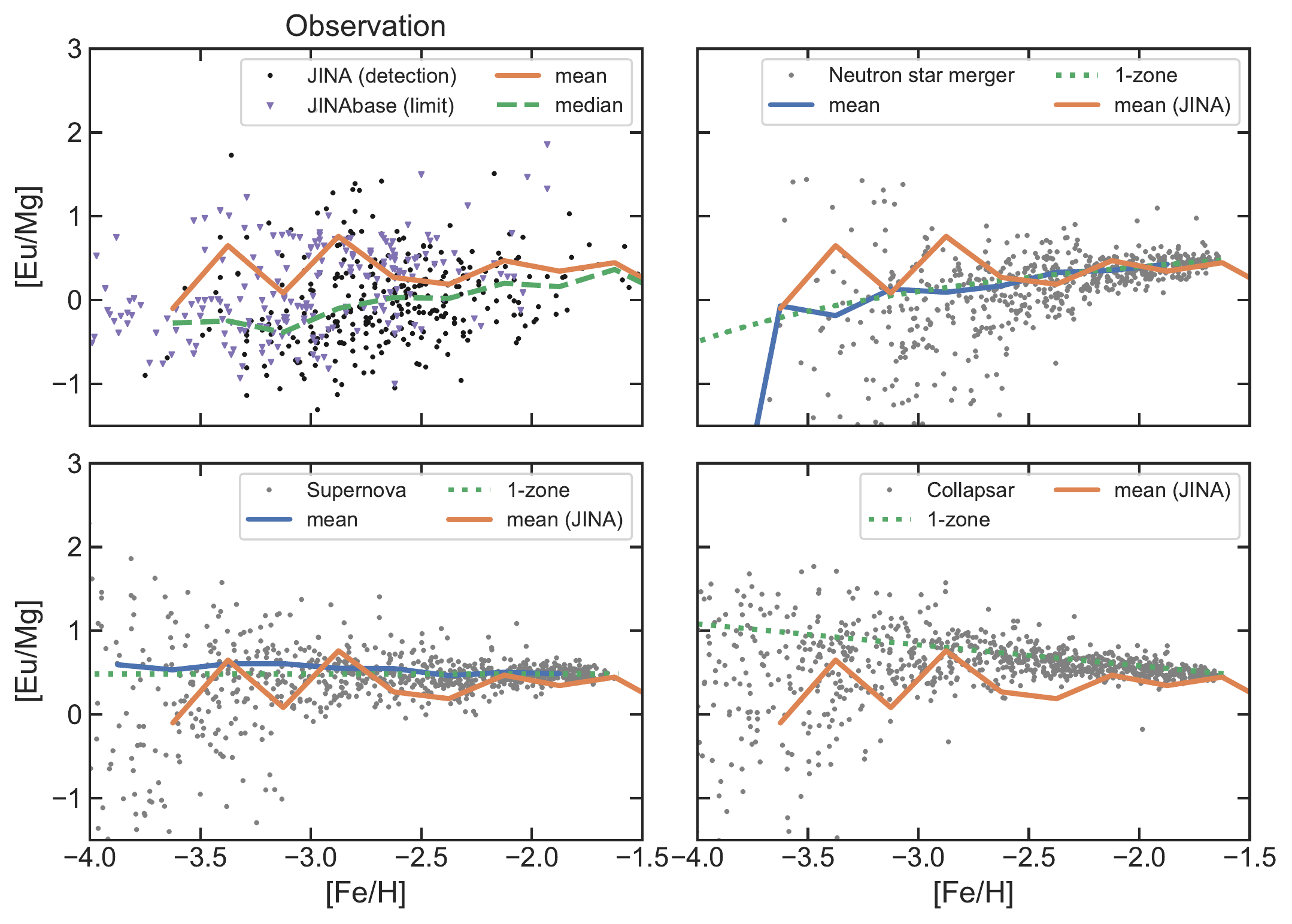}
    \caption{Same as Figure~\ref{fig:Model comparison} but with Eu. Note that  the upper limits are not included for the mean and median curves.}
    \label{fig:EuMg}
\end{figure}

In Figure~\ref{fig:EuMg} we show the analysis with Eu. The increasing trend we have discussed with Ba is not seen because there are a lot of upper limits and the observed population is significantly biased from the true distribution. However, the upper limits at [Fe/H]$\lesssim -3$ indicate a weak increasing trend in [Eu/Mg]

\bibliography{rprocess}{}
\bibliographystyle{aasjournal}


\listofchanges

\end{document}